\documentclass[a4paper,pra,twocolumn,showpacs,superscriptaddress,groupedaddress]{revtex4}
\usepackage{ae}
\usepackage[T1]{fontenc}
\usepackage[ansinew]{inputenc}
\usepackage{amsmath}
\usepackage{amssymb}
\usepackage[caption=false]{subfig}
\usepackage{multirow}
\usepackage{array}
\usepackage[]{graphicx}
\usepackage{wrapfig}

\usepackage{color}
\usepackage[colorlinks]{hyperref}
\usepackage{lscape}
\begin{document}
\title{On Strong Monogamy Conjecture in Four-Qubit System}
\author{Sumana Karmakar}
\email{sumanakarmakar88@gmail.com}
\affiliation{Department of Applied Mathematics, University of Calcutta, 92, A.P.C. Road, Kolkata-700009, India.}
\author{Ajoy Sen}
\email{ajoy.sn@gmail.com}
\affiliation{Department of Applied Mathematics, University of Calcutta, 92, A.P.C. Road, Kolkata-700009, India.}
\author{Amit Bhar}
\email{bhar.amit@yahoo.com}
\affiliation{Department of Mathematics, Jogesh Chandra Chaudhuri College, 30, Prince Anwar Shah Road, Kolkata-700033, India.}
\author{Debasis Sarkar}
\email{dsappmath@caluniv.ac.in,debasis1x@yahoo.co.in}
\affiliation{Department of Applied Mathematics, University of Calcutta, 92, A.P.C. Road, Kolkata-700009, India.}

\begin{abstract}
Monogamy is a defining feature of entanglement, having far reaching applications. Recently, Regula \textit{et.al.} in Phys. Rev. Lett. \textbf{113}, 110501(2014) have proposed a stronger version of monogamy relation for concurrence. We have extended the strong monogamy inequality for another entanglement measure, viz., negativity. In particular, we have concentrated on four-qubit system and provided a detail study on the status of strong monogamy on pure states. Further, we have analytically provided some classes of states for which negativity and squared negativity satisfy strong monogamy. Numerical evidences have also been shown in proper places. Our analysis also provides cases where strong monogamy is violated.
\end{abstract}
\date{\today}
\maketitle

\section{Introduction}
Many-body quantum system can show several interesting phenomena such as fractional Hall effect, superconductivity at high temperature in presence of strong correlation among the constituent bodies. Study of entanglement in multiparty scenario can provide explanations of such physical phenomena\cite{many body review}. Entanglement is an important physical resource behind several quantum information processing tasks like teleportation, dense coding \cite{quantum communication1,quantum communication2}, quantum computation\cite{quantum computation1,quantum computation2,quantum computation3}, quantum cryptography\cite{entanglement review} and even in some biological phenomena\cite{biology}. Hence, the task of characterizing and quantifying  entanglement has emerged as one of the prominent themes of quantum information theory.
Bi-partite entanglement is  well understood at least for qubit system but such characterization or classification in multipartite system is still very challenging.

Monogamy\cite{CKW,Tehral} is one of the key features of  multipartite entanglement. It puts restrictions on free sharing of entanglement among different parties and this particular restriction delineates quantumness from classicality via entanglement\cite{entanglement review}. Monogamy of entanglement is the key ingredient behind secure quantum cryptography\cite{monogamy cryptography} and has  important role in condensed matter physics such as n-representability problem for Fermions\cite{N-representability}. Thus, it is an important task to understand entanglement monogamy and its characterization in order to unveil the power of quantum entanglement in multipartite system. Monogamy of entanglement was first noted by Coffman, Kundu and Wootters\cite{CKW} in terms of squared concurrence through an inequality, often referred to as CKW-inequality. Concurrence is a well known measure of bipartite entanglement, introduced by Wootters\cite{concurrence1,concurrence2}. For a pure three-qubit state $|\Psi\rangle_{A_1A_2A_3}$, the CKW-inequality is given by
\begin{equation}\label{CKW}
C^2_{A_1|A_2A_3}\ge C^2_{A_1|A_2}+C^2_{A_1|A_3},
\end{equation}
where $A_1,A_2,A_3$ are three qubits with $A_1$ as focus qubit and the vertical bar indicates the bipartite splitting which enable one to compute concurrence $C$. Concurrence of a two-qubit state $\rho$  is defined as $C(\rho)= \max\{0,\lambda_1-\lambda_2-\lambda_3-\lambda_4\}$ in which $\lambda_1,\lambda_2,\lambda_3,\lambda_4$ are the square root of the eigenvalues of the  matrix $\rho(\sigma_y\otimes \sigma_y)\rho^*(\sigma_y\otimes \sigma_y)$ in decreasing order, $\sigma_y$ is the Pauli spin matrix and $\rho^*$ denotes the complex conjugate of $\rho$. $C_{A_1|A_2}$ and $C_{A_1|A_3}$ are  the concurrences of reduced density matrices $\rho_{A_1A_2}$ and $\rho_{A_1A_3}$ respectively of the state $\rho_{A_1A_2A_3}=|\Psi\rangle_{A_1A_2A_3}\langle \Psi|$. $C_{A_1|A_2A_3}$ is just the linear entropy of the subsystem $A_1$. This inequality has been extended to three-qubit mixed states in terms of generalized concurrences\cite{Generalized concurrence,CKW for mixed state}.
Coffman \textit{et.al.}\cite{CKW} have conjectured that the above monogamy relation (\ref{CKW}) can be extended to multipartite system. After several years, Osborne and Verstraete proved\cite{generalized CKW} this conjecture, i.e., a generalized version of inequality (\ref{CKW}) for $n$-qubit system
\begin{equation}\label{generalised CKW}
C^2_{A_1|A_2\ldots A_n}\ge C^2_{A_1|A_2}+\ldots+C^2_{A_1|A_n}.
\end{equation}
This relation suggests that entanglement between the subsystem $A_1$ with rest of the subsystems is greater than the sum of the pairwise entanglement of $A_1$ with each of the other $(n-1)$ parties. For $n>3$, the difference between left and right hand side of relation (\ref{generalised CKW}) gives a rough indicator of the amount of entanglement genuinely shared among $n$ qubits. In last few years, different attempts have been made to construct a proper  generalized version of monogamy inequality but those have not led to clear recipes to isolate the genuine $n$-partite entanglement.

Recently, Regula \textit{et.al.}\cite{SM inequality} have proposed a set of sharper version of monogamy inequality (\ref{generalised CKW}). For an $n$-qubit pure state $|\Psi\rangle$, this strong monogamy(in short SM) relation in terms of squared concurrence reads,
\begin{equation}\label{SM for tangle}
\begin{split}
\tau^{(1)}_{A_1|(A_2\ldots A_n)}\ge \sum_{j=2}^{n}\tau^{(2)}_{A_1|A_j}+\sum_{k>j=2}^{n}\left[\tau^{(3)}_{A_1|A_j|A_k}\right]^{\mu_3}+\\\sum_{l=2}^{n}\left[\tau^{(n-1)}_{A_1|A_2|\ldots|A_{l-1}|A_{l+1}|\ldots|A_n}\right]^{\mu_{n-1}},
\end{split}
\end{equation}
where, one-tangle $\tau^{(1)}_{A_1|(A_2...A_n)}$ is equivalent to the squared concurrence $C^2_{A_1|(A_2...A_n)}$ between the party $A_1$ and the rest of the system, two-tangle $\tau^{(2)}_{A_i|A_j}=C^2_{A_i|A_j}$, three-tangle $\tau^{(3)}_{A_1|A_j|A_k}=\tau^{(1)}_{A_1|(A_jA_k)}- \sum_{m=j,k}\tau^{(2)}_{A_1|A_m}$ and so on.  $\{\mu_m\}_{m=2}^{n-1}$ with $\mu_2=1$ is a sequence of rational exponents which can regulate the weight assigned to the different $m$-partite contributions. Thus, residual entanglement, namely, $n$-tangle ($\tau^{(n)}$) is defined as the difference between left and right hand side of  relation(\ref{SM for tangle}). This gives an indicator of the leftover entanglement, not distributed in all combinations with party $A_1$. SM inequality reduces to normal CKW inequality for three party system. Some examples, in support of this SM-inequality, have been shown for four-qubit pure states and this inequality has been conjectured to hold in multi-qubit pure systems too. Strong monogamy conjecture has been proved to hold good in $n$-qubit generalized W-class states in \cite{kim}.  Naturally, one can ask whether this conjecture is true for all qubit systems or there are cases of violation of this conjecture. This conjecture can also be extended to  other entanglement monotones rather than the concurrence. In this work, we have tried to answer this question by considering other measures of entanglement, viz., negativity and squared negativity. Like concurrence, negativity  is another useful measure of entanglement\cite{Negativity}. It is regarded as a quantitative version of Peres criterion of separability\cite{Peres}. Compared with  concurrence, the process of calculating  negativity is significantly simplified with respect to mixed states since it does not need the convex roof extension. In this paper, we will investigate similar types of SM-constraints, like relation (\ref{SM for tangle}), where instead of concurrence we will consider negativity and squared negativity as entanglement monotones. Particularly, we will provide classes of four-qubit pure states which satisfy our strong monogamy condition. We will also discuss the possibility of similar SM-conjecture in our case and provide support to our claim. Recently, Kim \textit{et.al.}\cite{kim2} have extended the concept of strong monogamy to square of convex roof extended negativity(SCREN) and showed that strong monogamy holds good even in higher dimensional system where original strong monogamy inequality fails. They have also showed that the original SM inequality can be obtained from their SCREN strong monogamy inequality.

We have organized our paper as follows: in section II, we set up the notations required for our investigation. Section III contains our results on four-qubit system. In section IV, we will discuss the status of strong monogamy for some interesting class of states and we conclude in section V.

\section{Strong Monogamy constraints in terms of Negativity}
On a quantitative level, \textit{negativity of entanglement} is considered as a valid, computable measure for bipartite entanglement. Negativity provides an alternative measure of mixed state entanglement that has the extremely rare property of being computable. The concept of negativity, originated from the Peres Criterion\cite{Peres}, states that  partial transpose of a density matrix, associated with a separable state, is still a valid density matrix, i.e., a positive semi-definite matrix. So, a state with a non-positive partial transpose must be entangled. Thus negativity of entanglement is based on the failure of the transpose operation to preserve positivity when acting on subsystems. Negativity was first introduced by Zyczkowski \textit{et. al.}\cite{Zyczkowski} and subsequently it was introduced as an entanglement measure by Vidal and Werner\cite{Negativity}.  Negativity $\mathcal{N}(\rho_{AB})$ of  a bipartite pure or mixed state $\rho$ of a $ d\otimes d'$ composite system  is defined as\cite{Negativity normalised},
\begin{equation}
\mathcal{N}(\rho_{AB})=\frac{\|\rho^{T_A}_{AB}\|_1-1}{d-1},
\end{equation}
where the trace norm $\|X \|_1$ is defined as $\|X\|_1=\text{tr}(\sqrt{XX^\dag})$. The quantity $\mathcal{N}(\rho_{AB})$ is equivalent to twice the sum of the absolute values of negative eigenvalues of $\rho^{T_A}_{AB}$, where $\rho^{T_A}_{AB}$ is the partial transpose of the density matrix $\rho_{AB}$ with respect to the subsystem $A$. This additive measure is an entanglement monotone, invariant under local unitary and considered as an important measure of entanglement.\\

Negativity, in general, does not satisfy the monogamy relation. However, as given by He and Vidal\cite{vidal}, it can satisfy monogamy relation in some settings as provided by disentangling theorem.  Ou and Fun\cite{CKW for negativity} proved the CKW-type monogamy inequality in terms of squared negativity as,
\begin{equation}\label{CKW for negativity}
\mathcal{N}^2_{A_1|A_2A_3}\ge \mathcal{N}^2_{A_1|A_2}+\mathcal{N}^2_{A_1|A_3}.
\end{equation}
The difference between the two sides of relation (\ref{CKW for negativity}) can be interpreted as the residual entanglement which is a quantifier of entanglement, genuinely shared among three parties. This quantity may also be called as monogamy score.  We can define this residual entanglement, namely, three-$\pi$ entanglement as,
\begin{equation}\label{three pi}
 \pi^{(3)}_{A_1|A_2|A_3}=\mathcal{N}^2_{A_1|A_2A_3}-\mathcal{N}^2_{A_1|A_2}-\mathcal{N}^2_{A_1|A_3},
\end{equation}
which is used to characterize three way entanglement of a state. Interestingly, unlike three-tangle, this quantity depends on focus qubit (here $A_1$), i.e., $\mathcal{N}^2_{A_1|A_2A_3}-\mathcal{N}^2_{A_1|A_2}-\mathcal{N}^2_{A_1|A_3}\ne\mathcal{N}^2_{A_2|A_1A_3}-\mathcal{N}^2_{A_2|A_1}-\mathcal{N}^2_{A_2|A_3}\ne\mathcal{N}^2_{A_3|A_1A_2}-\mathcal{N}^2_{A_3|A_1}-\mathcal{N}^2_{A_3|A_2}.$ This indicates that the residual entanglement varies under permutations of the parties. This feature entails an inherent asymmetry of entanglement sharing in multiparty system and it seems obvious rather than symmetric sharing.\\

For $n$-qubit pure state $|\Psi\rangle$, SM-inequality, corresponding to negativity, takes the following form,
\begin{widetext}
\begin{equation}\label{SM inequality for negativity}
\begin{split}
\delta^{(1)}_{A_1|A_2\ldots A_n}(|\Psi\rangle)\ge \sum_{j=2}^{n}\delta^{(2)}_{A_1|A_j}(|\Psi\rangle)+\sum_{k>j=2}^{n}\left[\delta^{(3)}_{A_1|A_j|A_k}(|\Psi\rangle)\right]^{\mu_3}+
\ldots+\sum_{l=2}^{n}\left[\delta^{(n-1)}_{A_1|A_2|\ldots|A_{l-1}|A_{l+1}|\ldots|A_n}(|\Psi\rangle)\right]^{\mu_{n-1}},
\end{split}
\end{equation}
\end{widetext}
where, $\delta^{(1)}_{A_1|A_2\ldots A_n}:=\mathcal{N}_{A_1|A_2\ldots A_n}$, $\delta^{(2)}_{A_i|A_j}:=\mathcal{N}_{A_i|A_j}$, $\delta^{(3)}_{A_1|A_j|A_k}:=\delta^{(1)}_{A_1|(A_jA_k)}- \sum_{m=j,k}\delta^{(2)}_{A_1|A_m}$ and so on.
Similarly, for squared negativity it takes the following form,
\begin{widetext}
\begin{equation}\label{SM inequality for squared negativity}
\begin{split}
\pi^{(1)}_{A_1|A_2...A_n}(|\Psi\rangle)\ge \sum_{j=2}^{n}\pi^{(2)}_{A_1|A_j}(|\Psi\rangle)+\sum_{k>j=2}^{n}\left[\pi^{(3)}_{A_1|A_j|A_k}(|\Psi\rangle)\right]^{\mu_3}+
\ldots+\sum_{l=2}^{n}\left[\pi^{(n-1)}_{A_1|A_2|\ldots|A_{l-1}|A_{l+1}|\ldots|A_n}(|\Psi\rangle)\right]^{\mu_{n-1}},
\end{split}
\end{equation}
\end{widetext}
where, $\pi^{(1)}_{A_1|A_2\ldots A_n}:= \mathcal{N}^2_{A_1|A_2...A_n}$, $\pi^{(2)}_{A_i|A_j}:=\mathcal{N}^2_{A_i|A_j}$ and $\pi^{(3)}_{A_1|A_j|A_k}:=\pi^{(1)}_{A_1|(A_jA_k)}- \sum_{m=j,k}\pi^{(2)}_{A_1|A_m},$ etc. In both the cases, $\{\mu_{m}\}_{m=2}^{n-1}$ is a sequence of rational  exponents with $\mu_2\equiv1$, which can regulate the weight assigned to the different $m$-partite contributions. Then, we define the n-partite pure state residual entanglement or monogamy score, namely $n$-delta( $\delta^{(n)}$) and $n$-pi( $\pi^{(n)}$), as the difference between left and right hand side of relation (\ref{SM inequality for negativity}) and (\ref{SM inequality for squared negativity}) respectively, i.e.,
\begin{widetext}
\begin{equation}\label{n-delta}
\begin{split}
\delta^{(n)}_{A_1|A_2|\ldots|A_n}(|\Psi\rangle) : = \delta^{(1)}_{A_1|A_2\ldots A_n}(|\Psi\rangle)- \sum_{j=2}^{n}\delta^{(2)}_{A_1|A_j}(|\Psi\rangle)-\sum_{k>j=2}^{n}\left[\delta^{(3)}_{A_1|A_j|A_k}(|\Psi\rangle)\right]^{\mu_3}-\ldots \\
-\sum_{l=2}^{n}\left[\delta^{(n-1)}_{A_1|A_2|\ldots|A_{l-1}|A_{l+1}|\ldots|A_n}(|\Psi\rangle)\right]^{\mu_{n-1}}
\end{split}
\end{equation}
and
\begin{equation}\label{n-pi}
\begin{split}
\pi^{(n)}_{A_1|A_2|\ldots|A_n}(|\Psi\rangle) := \pi^{(1)}_{A_1|A_2\ldots A_n}(|\Psi\rangle)- \sum_{j=2}^{n}\pi^{(2)}_{A_1|A_j}(|\Psi\rangle)-\sum_{k>j=2}^{n}\left[\pi^{(3)}_{A_1|A_j|A_k}(|\Psi\rangle)\right]^{\mu_3}-\ldots \\
 -\sum_{l=2}^{n}\left[\pi^{(n-1)}_{A_1|A_2|\ldots|A_{l-1}|A_{l+1}|\ldots|A_n}(|\Psi\rangle)\right]^{\mu_{n-1}}.
\end{split}
 \end{equation}
\end{widetext}
We remark that the above SM-inequalities hold only when all the terms $\delta^{(m)}$ and $\pi^{(m)}$, appearing in relation (\ref{SM inequality for negativity}) and (\ref{SM inequality for squared negativity}) respectively, are non-negative.  Any non trivial rational sequence $\{\mu_m\}$ with $\mu_2=1$ implies  a particular SM-inequality. The validity of SM-inequality for a particular $\{\mu_m^*\}$ implies the validity of relation (\ref{SM inequality for negativity}) for all exponents $\mu_m^*\le\mu_m$. Due to this reason, one should attempt to choose $\{\mu_m\}$ as small as possible and thus  $\{\mu_m\}=1$ ( for all $m$) is the minimal choice. In our work, we will consider four-qubit system and hence strong monogamy constraints reduce to  $\delta^{(4)}\ge 0$ and $\pi^{(4)}\ge 0$. Here, we will only consider the class of states where negativity promises to be monogamous.

\section{Strong Monogamy in Four-Qubit Generic Class}
Analytic computation of the expressions of strong monogamy relation for negativity is a formidable task in multiparty system, in general. Here, we will explore the SM-inequality for negativity in four-qubit system. Four-qubit pure states can be classified into different inequivalent ways\cite{nine class}. Among them, the generic class (denoted by $\mathcal{A}$) is an important class of pure states and it is defined as\cite{nine class,generic class}
\begin{equation}
\begin{split}
\mathcal{A}=\{z_1u_1+z_2u_2+z_3u_3+z_4u_4 \quad|\quad z_1,z_2,z_3,z_4\in \mathbb{C}\\ \quad \text{and} \quad\sum_{i=1}^4|z_i|^2=1 \},
\end{split}
\end{equation}
with $u_1\equiv|\Phi^+\rangle|\Phi^+\rangle$, $u_2\equiv|\Phi^-\rangle|\Phi^-\rangle$, $u_3\equiv|\Psi^+\rangle|\Psi^+\rangle$, $u_4\equiv|\Psi^-\rangle|\Psi^-\rangle$, $|\Phi^{\pm}\rangle=\frac{|00\rangle\pm|11\rangle}{\sqrt{2}}$ and  $|\Psi^{\pm}\rangle=\frac{|01\rangle\pm|10\rangle}{\sqrt{2}}$. This class is called generic because under the action of SLOCC (stochastic local operation and classical communication) operation, this class is dense in the space of four-qubits. This class even contains uncountable SLOCC inequivalent subclasses. We will also consider two subclasses $\mathcal{B}$ and $\mathcal{C}$ of the class $\mathcal{A}$:
\begin{equation}
\mathcal{B}=\{|\Phi \rangle_{ABCD}\in \mathcal{A}: z_1=z_2, z_3=z_4\},
\end{equation}
\begin{equation}
\begin{split}
\mathcal{C}=\{z_1u_1+z_2u_2+z_3u_3+z_4u_4 \quad|\quad z_1,z_2,z_3,z_4\in \mathbb{R}\\\text{and} \quad\sum_{i=1}^4 z_i^2=1 \}.
\end{split}
\end{equation}

\begin{figure}[htb]
\subfloat[]{\includegraphics[scale=0.5]{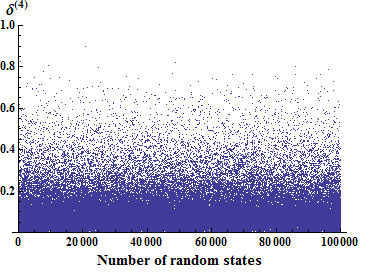}\label{n2}}\\
\subfloat[]{\includegraphics[scale=0.5]{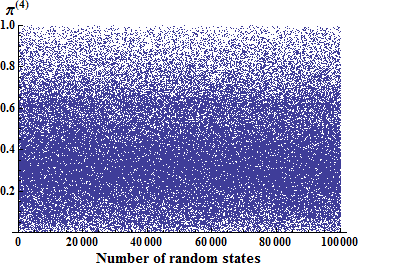}\label{sn2}}\\
\subfloat[]{\includegraphics[scale=0.7]{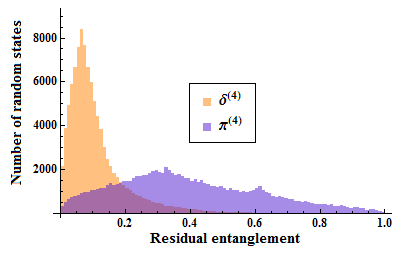}\label{hist}}
\caption{(Color online)Numerical simulation of $10^5$ random states from the class $\mathcal{C}$ showing strong evidence that $\delta^{(4)}$ and $\pi^{(4)}$ are non-negative in all the cases. In the FIG.(\ref{n2}) we have considered $\mu_3=1.5$ and we have considered $\mu_3=1$ in FIG.(\ref{sn2}). In the FIG.(\ref{hist}) we have shown the superposed histograms of strong monogamy scores $\delta^{(4)}$ and $\pi^{(4)}$  corresponding to negativity and squared negativity.}
\label{SM of class C}
\end{figure}

All the terms $\mathcal{\delta}^{(m)}$ and $\mathcal{\pi}^{(m)}$, in relations (\ref{SM inequality for negativity}) and (\ref{SM inequality for squared negativity}) respectively, are non-negative corresponding to the subclass $\mathcal{B}$. Hence, strong monogamy  relations (\ref{SM inequality for negativity}) and (\ref{SM inequality for squared negativity}) are well defined for this class of states and both SM-inequalities (relations (\ref{SM inequality for negativity}) and (\ref{SM inequality for squared negativity})) are satisfied by  most of the states of class $\mathcal{B}$. Strong monogamy is violated whenever two complex parameters are orthogonal (Refer Appendix \ref{class B}). For class $\mathcal{C}$,  whenever $\delta^{(3)}\ge 0$, our numerical evidence suggests that this class, possibly, satisfies SM-inequality (\ref{SM inequality for negativity}), depending on the proper choice of the sequence $\{\mu_m\}$. Similar conclusion holds(relation (\ref{SM inequality for squared negativity})) for squared negativity whenever  $\pi^{(3)}\ge 0$. We have performed numerical simulation with $10^5$ random states from this class $\mathcal{C}$, satisfying the constraints $\delta^{(3)}\ge 0$ or $\pi^{(3)}\ge 0$ respectively,  and plotted $\delta^{(4)}$ and $\pi^{(4)}$ for suitable weight value (refer FIG. \ref{SM of class C}). The result provides strong evidence that SM-inequalities are satisfied by both the subclasses. In the next section, we will discuss the status of strong monogamy for some important subclasses of four-qubit pure states.

\section{Strong Monogamy in Some Particular Class of states}
\textbf{Cluster states:} Cluster states are typically multipartite entangled states and they are utilized in quantum error-correcting codes\cite{cluster error correction} and testing of quantum nonlocality\cite{cluster nonlocality}. Moreover, they are also a universal resource in one-way quantum computation\cite{quantum computation2}. Four-qubit cluster states can be written as\cite{cluster class}
\begin{equation}\label{cluster state}
|\Psi^c\rangle_{A_1A_2A_3A_4}=a|0000\rangle+b|0011\rangle+c|1100\rangle-d|1111\rangle,
\end{equation}
where $a,b,c,d\in \mathbb{C}$ with $|a|^2+|b|^2+|c|^2+|d|^2=1.$
This class of states satisfy (refer Appendix \ref{cluster}) the SM-inequalities (\ref{SM inequality for negativity}) and (\ref{SM inequality for squared negativity}) under the  constraints $\delta^{(3)}\ge 0$ and $\pi^{(3)}\ge 0$ respectively with  minimal choice of $\{\mu_m\}$ (i.e., $\mu_m=1$, $\forall m$).\\

\textbf{Dicke states:} $n$-qubit Dicke state with $ k $-excitations is  given by
\begin{equation}\label{dicke state}
|S(n,k)\rangle\equiv\sqrt{\frac{k!(n-k)!}{n!}}\sum_{\text{permutations}}|0\rangle^{\otimes(n-k)}|1\rangle^{\otimes k},
\end{equation}
where the summation is over all possible permutations of the product state having $k$ qubits in the excited state $|1\rangle$ and $(n-k)$ qubits in the ground state $|0 \rangle $. Here we will consider four-qubit Dicke states. Among these states, $ |S(4,0)\rangle $ and $ |S(4,4)\rangle $ are not entangled. $ |S(4,2)\rangle $ is entangled but for this state $\delta^{(4)}$ does not exist because this state does not satisfy normal monogamy condition w.r.t. negativity. However, this state satisfies strongly monogamy in case of squared negativity with $\mu_m=1$, $\forall m$. The states  $ |S(4,1)\rangle $ and  $ |S(4,3)\rangle $ are, in fact, W and anti W-state (denoted by $\tilde{W}$ ) respectively. Both the states satisfy strong monogamy relation (refer Appendix \ref{dicke}).\\

\textbf{Superposition of the $W$ and $\tilde{W}$ states:} We, now, consider superposition of the $W$ and $\tilde{W}$ states as
\begin{equation}\label{superposition state of W and Wc}
|W\tilde{W}(s,\phi)\rangle\equiv \sqrt{s}|W\rangle+\sqrt{1-s}e^{i \phi}|\tilde{W}\rangle,
\end{equation}
where $0<s<1$ and $\phi\in [0,2\pi].$
This class satisfies both the SM-inequalities. We have found that $\delta^{(4)}> 0$ for $\mu_3=1.04$ and $\pi^{(4)}> 0$ for $\mu_3= 1$ (refer Appendix \ref{super}).\\

\textbf{Generalized GHZ state:} Four-qubit generalized GHZ state can be written in terms of superposition of particular Dicke states as
\begin{equation}\label{GGHZ}
 |GGHZ\rangle=z_1|S(4,0)\rangle+z_2|S(4,4)\rangle,
\end{equation}
 where $z_1$, $z_2\in\mathbb{C}$ and $|z_1|^2+|z_2|^2=1$. For these states, simple calculation yields $\delta^{(4)}=2|z_1z_2|$ and $\pi^{(4)}=4|z_1z_2|^2$. Hence, $\delta^{(4)}> 0$ and $\pi^{(4)}> 0$ for $\mu_m= 1$, $\forall m$. Thus, this class of states are strongly monogamous.\\

\begin{figure}[h]
    \subfloat[]{\includegraphics[scale=0.5]{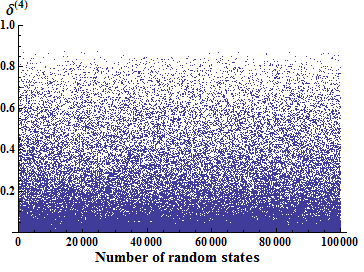}\label{w-0-superposition}}\\ \subfloat[]{\includegraphics[scale=0.5]{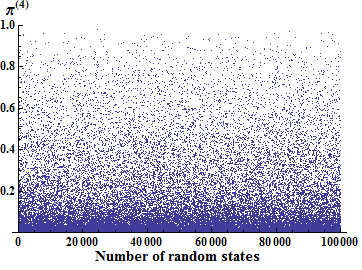}\label{w-0-superposition-sqr}}\\ \subfloat[]{\includegraphics[scale=0.7]{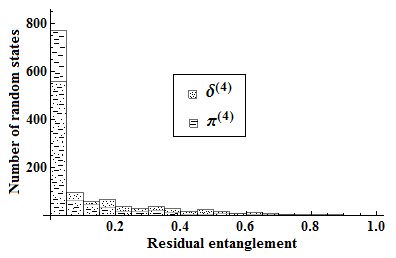}\label{hist2}}\\
     \caption{(Color online) Fig. (\ref{w-0-superposition}) and (\ref{w-0-superposition-sqr}) describe the nature of $\delta^{(4)}$ and $\pi^{(4)}$ for superposed state $|\Phi_p\rangle$ in Eqn.(\ref{superposition of GW and product state 0}). $\delta^{(4)}$ and $\pi^{(4)}$ are non-negative for $\mu_3= 3$ and $\mu_3 = 2.5$ respectively. The superposed histograms of $\delta^{(4)}$ and $\pi^{(4)}$ (instead of $10^5$, $10^3$ simulations are shown for clarity) are showed in Fig. (\ref{hist2}). The statistical nature of numerical data shows sharp peak near origin and hence in these  cases strong monogamy is satisfied by very small margin.}
     \label{figure for w and product state 0 superposition}
 \end{figure}

\textbf{Superposition of generalized $W$ and  $|0\rangle^{\otimes 4}$:} Superposition of four-qubit generalized $W$-class state and the pure product state $|0\rangle^{\otimes 4}$ can be written as
\begin{equation}\label{superposition of GW and product state 0}
|\Phi_p\rangle=\sqrt{p}|GW\rangle+\sqrt{1-p}|0\rangle^{\otimes 4},
\end{equation}
where $|GW\rangle = a_1|0001\rangle+a_2|0010\rangle+a_3|0100\rangle+a_4|1000\rangle$ is the four-qubit generalized $W$-class state with the normalization condition $\sum_{1}^{4}|a_i|^2=1$ and $0<p<1.$
We have numerical results corresponding to this class as we have plotted $\delta^{(4)}$ and $\pi^{(4)}$ for $10^5$ random states from the class. Our numerical evidence suggests that  strong monogamy  relations (\ref{SM inequality for negativity}) and (\ref{SM inequality for squared negativity}) are well defined for this class of states and the inequalities  hold good for proper choices of weight values. In fact, we get  $\delta^{(4)}\ge 0$ for $\mu_3=3$ and $\pi^{(4)}\ge 0$ for $\mu_3= 2.5$ (refer FIG. \ref{figure for w and product state 0 superposition}).

\begin{figure}[htb]
   \subfloat[]{\includegraphics[scale=0.6]{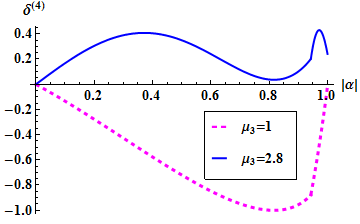}\label{w-1-superposition}}\\
   \subfloat[]{\includegraphics[scale=0.6]{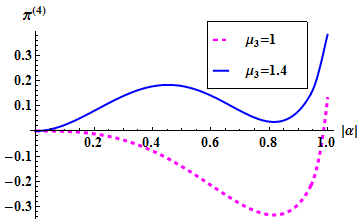}\label{w-1-superposition-sqr}}\\
    \caption{(Color online) Fig. (\ref{w-1-superposition}) and (\ref{w-1-superposition-sqr}) describe the nature of $\delta^{(4)}$ and $\pi^{(4)}$ for superposed state $|\Phi_{\alpha,\beta}\rangle$ in Eqn.(\ref{superposition of W and product state 1}). $\delta^{(4)}$ is non-negative for $\mu_3= 2.8$ (solid blue line) but negative for $\mu_3= 1$ (dashed magenta line). Similarly, $\pi^{(4)}$ is non-negative for $\mu_3= 1.4$ (solid  blue line) but negative and $\mu_3= 1$ (dashed magenta line).}
    \label{figure for w and product state 1 superposition}
\end{figure}

\textbf{Superposition of $W$ and  $|1\rangle^{\otimes 4}$:} Any superposition of four-qubit $W$ state and the pure product state $|1\rangle^{\otimes 4}$ can be written as
\begin{equation}\label{superposition of W and product state 1}
|\Phi_{\alpha,\beta}\rangle=\alpha|W\rangle+\beta|1\rangle^{\otimes 4},
\end{equation}
where $\alpha,\beta\in\mathbb{C}$ and $|\alpha|^2+|\beta|^2=1 $. SM-inequalities  hold good for this class with proper choice of weight values. In fact, we get  $\delta^{(4)}> 0$ for $\mu_3=2.8$ and $\pi^{(4)}\ge 0$ for $\mu_3= 1.4 $ (refer Appendix \ref{superw}). In FIG. \ref{figure for w and product state 1 superposition}, we have shown the dependency of monogamy on the chosen weight values.

\section{Conclusion}
In this work, we  have analyzed  strong monogamy inequality in terms of negativity and squared negativity. We have found a subclass of four-qubit generic class of states for which strong monogamy is satisfied by both negativity  and squared negativity. We have also investigated the status of strong monogamy on different classes of four-qubit pure states. We have noticed that for the proper choice of weight sequence most of these classes satisfy strong monogamy. We have also found few classes of states for which negativity and squared negativity are not strongly monogamous. Numerical evidence of strong monogamy has also been presented for a particular subclass of generic class  and for the superposed class of four qubit generalised $W$ and ground state, with the proper choice of the weight values.  The results suggest that both negativity and square negativity satisfy strong monogamy for these classes with proper weight function. Our result clearly shows the boundary for strong monogamy inequality for four qubit systems and hence also for SCERN. Future directions may include the study of overcoming our limitations in computationally difficult areas and the study of strong monogamy relations for other entanglement monotones, like, squashed entanglement\cite{squased1,squashed2}.
We hope our results will help to improve the current understanding of entanglement sharing through its monogamy structure, which will be useful to characterize many-body phenomena.

\begin{acknowledgements}
	The author S. Karmakar acknowledges the financial support from University Grants Commission, New Delhi, India. The author D. Sarkar also acknowledges DST SERB for financial support.
\end{acknowledgements}

\appendix

\section{Strong Monogamy in class $\mathcal{B}$}\label{class B}
Let us consider a pure state $|\Psi\rangle\in\mathcal{B}$. Using some simple algebra, we can obtain,
\begin{equation}
\begin{split}
&\mathcal{N}_{A_1|A_2A_3A_4}=1,\\
&\mathcal{N}_{A_1|A_2A_3}=\mathcal{N}_{A_1|A_3A_4}=4|z_1z_3|,\\
&\mathcal{N}_{A_1|A_3}=4|Re(z_1^*z_3)|,\\
&\mathcal{N}_{A_1|A_2A_4}=\mathcal{N}_{A_1|A_2}=\mathcal{N}_{A_1|A_4}=0.
\end{split}
\end{equation}
Using the expressions of $\delta^{(m)}$'s and $\pi^{(m)}$'s in the relations (\ref{n-delta}) and (\ref{n-pi}), we have,
\begin{equation}\label{n4}
\begin{split}
\delta^{(4)}_{q_1|q_2|q_3|q_4}&=\delta^{(1)}_{q_1|q_2q_3q_4}-\sum_{j=2}^4\delta^{(2)}_{q_1|q_j}-\sum_{k>j=2}^4\left[\delta^{(3)}_{q_1|q_j|q_k}\right]^{\mu_3}\\
&=1-2^2|Re(z_1^*z_3)|-2^{2\mu_3+1}(|z_1z_3|\\&-|Re(z_1^*z_3)|)^{\mu_3}
\end{split}
\end{equation}\
and
\begin{equation}\label{n5}
\begin{split}
\pi^{(4)}_{q_1|q_2|q_3|q_4}&=\pi^{(1)}_{q_1|q_2q_3q_4}-\sum_{j=2}^4\pi^{(2)}_{q_1|q_j}-\sum_{k>j=2}^4\left[\pi^{(3)}_{q_1|q_j|q_k}\right]^{\mu_3}\\
&=1-2^4|Re(z_1^*z_3)|^2-2^{4\mu_3+1}(|z_1z_3|^2\\&-|Re(z_1^*z_3)|^2)^{\mu_3}.
\end{split}
\end{equation}
Substituting $z_1=r_1(\cos\alpha+\imath\sin\alpha)$ and $z_3=r_3(\cos\beta+\imath\sin\beta)$ with $\alpha, \beta\in[0,2\pi]$ and $r_1^2+r_3^2=\frac{1}{2}$ in Eqns.(\ref{n4}) and (\ref{n5}), we get,
\begin{equation}
\begin{split}
\delta^{(4)}_{A_1|A_2|A_3|A_4}=1-4r_1r_3|\cos(\alpha-\beta)|-\\2(4r_1r_3)^{\mu_3}(1-|\cos(\alpha-\beta)|)^{\mu_3},
\end{split}
\end{equation}
and
\begin{equation}
\begin{split}
\pi^{(4)}_{A_1|A_2|A_3|A_4}=1-(4r_1r_3)^2|\cos(\alpha-\beta)|^2-\\2(4r_1r_3)^{2\mu_3}(1-|\cos(\alpha-\beta)|^2)^{\mu_3}.
\end{split}
\end{equation}
It is clear that for high value of $\mu_3$, both the terms $(1-|\cos(\alpha-\beta)|)^{\mu_3} $ and $(1-|\cos(\alpha-\beta)|^2)^{\mu_3}$ tend to zero when $\alpha\ne\beta+(2n+1)\frac{\pi}{2}$, $n$ being an integer and therefore $\delta^{(4)}$ and $\pi^{(4)}$ become non-negative.

\section{Cluster states}\label{cluster}
It can be easily shown for any four-qubit cluster state $a|0000\rangle+b|0011\rangle+c|1100\rangle-d|1111\rangle,$
\begin{equation}
\begin{split}
&N_{A_1|A_2A_3A_4}=2\sqrt{(|a|^2+|b|^2)(|c|^2+|d|^2)},\\
&N_{A_1|A_2A_3}=4|ac^*-bd^*|,\\
&N_{A_1|A_2}=2|ac^*-bd^*|,\\
&N_{A_1|A_2A_4}=N_{A_1|A_3A_4}=N_{A_1|A_3}=N_{A_1|A_4}=0.
\end{split}
\end{equation}
Whenever $ac^*=bd^*$, all the residual terms $\delta^{(3)}$ and $\pi^{(3)}$ become zero. Substituting these into the relations (\ref{SM inequality for negativity}) and (\ref{SM inequality for squared negativity}), we get
\begin{equation}
\begin{split}
\delta^{(4)}_{A_1|A_2|A_3|A_4}=2\sqrt{(|a|^2+|b|^2)(|c|^2+|d|^2)}> 0,
\end{split}
\end{equation}
\begin{equation}
\begin{split}
\pi^{(4)}_{A_1|A_2|A_3|A_4}=4(|a|^2+|b|^2)(|c|^2+|d|^2)> 0.
\end{split}
\end{equation}

\section{Dicke states}\label{dicke}
Four-qubit Dicke state, with $1$ and $3$ excitation(s) respectively, can be written as,
\begin{eqnarray*}
|S(4,1)\rangle\equiv|W\rangle &=&(|0001\rangle+|0010\rangle+|0100\rangle+|1000\rangle)/2, \\
|S(4,3)\rangle\equiv |\tilde{W}\rangle &=&(|0111\rangle+|1011\rangle+|1101\rangle+|1110\rangle)/2.
\end{eqnarray*}
These states are connected by local unitary and hence they have same entanglement content. The results also reveal the similar feature. For these states we can easily derive,
\begin{equation}
\begin{split}
\delta^{(4)}=2(3+\sqrt{3}-3\sqrt{2})\frac{k!(n-k)!}{n!}-3(6-4\sqrt{2})^{\mu_3}\\\left(\frac{k!(n-k)!}{n!}\right)^{\mu_3},
\end{split}
\end{equation}
\begin{equation}
\begin{split}
\pi^{(4)}= 24(\sqrt{2}-1)(\frac{k!(n-k)!}{n!})^2-3(16\sqrt{2}-20)^{\mu_3}\\\left(\frac{k!(n-k)!}{n!}\right)^{2\mu_3}.
\end{split}
\end{equation}
Numerical solution suggests, $\delta^{(4)}> 0$ for $\mu_3\ge 1.02053$  and $\pi^{(4)}> 0$ for $\mu_3= 1$.\\

\section{Superposition of the $W$ and $\tilde{W}$}\label{super}
It is easy to see that entanglement of this class is independent of the phase factor $\phi$, because the transformation $\{|0\rangle,|1\rangle\}\rightarrow \{|0\rangle,e^{-i \phi}|1\rangle\}$ induces $|W\tilde{W}(s,\phi)\rangle \rightarrow e^{-i\phi}|W\tilde{W}(s,0)\rangle$. For these states,
\begin{equation}
\begin{split}
\delta^{(1)}&=\frac{1}{2}\sqrt{3+4s-4s^2},\\
\delta^{(2)}&=\frac{1}{2}(\sqrt{1+(1-2s)^2}-1),\\
\delta^{(3)}&=1+\frac{1}{2}|1-2s|-\sqrt{1+(1-2s)^2}
\end{split}
\end{equation}
and
\begin{equation}
\begin{split}
\pi^{(1)}&=\frac{1}{4}(3+4s-4s^2),\\
\pi^{(2)}&=\frac{3}{4}-s+s^2-\frac{1}{2}\sqrt{1+(1-2s)^2},\\
\pi^{(3)}&=-\frac{5}{4}+s-s^2+\sqrt{1+(1-2s)^2}.
\end{split}
\end{equation}
Hence
\begin{equation}
\begin{split}
\delta^{(4)}=\frac{1}{2}\sqrt{3+4s-4s^2}-\frac{3}{2}(\sqrt{1+(1-2s)^2}-1)- \\3(1+\frac{1}{2}|1-2s|-\sqrt{1+(1-2s)^2})^{\mu_3},
\end{split}
\end{equation}
\begin{equation}
\begin{split}
\pi^{(4)}=-\frac{3}{2}+4s-4s^2+\frac{3}{2}\sqrt{1+(1-2s)^2}-\\ 3\left(-\frac{5}{4}+s-s^2+\sqrt{1+(1-2s)^2}\right)^{\mu_3}.\\
\end{split}
\end{equation}

\section{Superposition of W state and $|1\rangle^{\otimes 4}$}\label{superw}
Like previous cases, using simple algebra, we obtain,
\begin{equation}
\begin{split}
\delta^{(1)}&=x,\\
\delta^{(2)}&=
\begin{cases}
0 &\mbox{if }|\alpha|\le\frac{2\sqrt{2}}{3}  \\
 z & \mbox{if } |\alpha|\ge\frac{2\sqrt{2}}{3}
\end{cases},\\
\delta^{(3)}&=
\begin{cases}
 y &\mbox{if }|\alpha|\le\frac{2\sqrt{2}}{3}  \\
 y- 2z & \mbox{if } |\alpha|\ge\frac{2\sqrt{2}}{3}
\end{cases}.
\end{split}
\end{equation}
Hence,
\begin{equation}
\begin{split}
\delta^{(4)}=
\begin{cases}
  x-3y^{\mu_3} &\mbox{if }|\alpha|\le\frac{2\sqrt{2}}{3}  \\
   x-3z-3(y-2z)^{\mu_3} & \mbox{if } |\alpha|\ge\frac{2\sqrt{2}}{3}
\end{cases}.
\end{split}
\end{equation}
Again,
\begin{equation}
\begin{split}
\pi^{(1)}&=x^2,\\
\pi^{(2)}&=
\begin{cases}
    0 &\mbox{if }|\alpha|\le\frac{2\sqrt{2}}{3}  \\
     z^2 & \mbox{if } |\alpha|\ge\frac{2\sqrt{2}}{3}
\end{cases},\\
\pi^{(3)}&=
\begin{cases}
     y^2 &\mbox{if }|\alpha|\le\frac{2\sqrt{2}}{3}  \\
     y^2- 2z^2 & \mbox{if } |\alpha|\ge\frac{2\sqrt{2}}{3}
\end{cases}.
\end{split}
\end{equation}
Hence,
\begin{equation}
\begin{split}
\pi^{(4)}=
\begin{cases}
      x^2-3y^{2\mu_3} &\mbox{if }|\alpha|\le\frac{2\sqrt{2}}{3}  \\
       x-3z^2-3(y^2-2z^2)^{\mu_3} & \mbox{if } |\alpha|\ge\frac{2\sqrt{2}}{3}
\end{cases},
\end{split}
\end{equation}
where $x=\frac{\sqrt{3}}{2}|\alpha|\sqrt{4-3|\alpha|^2}$, $y=\frac{|\alpha|}{4}(|\alpha|+\sqrt{16-15|\alpha|^2})$, $z=\frac{1}{2}(\sqrt{10|\alpha|^4-12|\alpha|^2+4}+|\beta|^2-2)$.
\end{document}